\documentclass[english]{aa}

\usepackage[varg]{txfonts}
\usepackage{natbib}
\usepackage{graphicx}
\usepackage{amsmath}  
\usepackage[rightcaption]{sidecap}
\usepackage[colorlinks=true,
    linkcolor=blue, citecolor=blue, filecolor=blue, urlcolor=blue]{hyperref}

\begin{document}

\title{Characterizing Jupiter’s interior using machine learning reveals four key structures}
\author{M. Ziv\inst{1}, E. Galanti\inst{1}, S. Howard\inst{2}, T. Guillot\inst{3},\and Y. Kaspi\inst{1}}
\institute{Department of Earth and Planetary Sciences, Weizmann Institute of Science, Rehovot 76100, Israel\\
\email{maayan.ziv@weizmann.ac.il}\and Institut für Astrophysik, Universität Zürich, Winterthurerstr. 190, 8057 Zürich, Switzerland \and Université Côte d'Azur, Observatoire de la Côte d'Azur, CNRS, Laboratoire Lagrange, France}
\date{Received 26 September 2024 / Accepted 12 November 2024}
\abstract
{The internal structure of Jupiter is constrained by the precise gravity field measurements by NASA's Juno mission, atmospheric data from the Galileo entry probe, and Voyager radio occultations. Not only are these observations few compared to the possible interior setups and their multiple controlling parameters, but they remain challenging to reconcile. As a complex, multidimensional problem, characterizing typical structures can help simplify the modeling process.}
{We explored the plausible range of Jupiter's interior structures using a coupled interior and wind model, identifying key structures and effective parameters to simplify its multidimensional representation.}
{We used NeuralCMS, a deep learning model based on the accurate concentric Maclaurin spheroid (CMS) method, coupled with a fully consistent wind model to efficiently explore a wide range of interior models without prior assumptions. We then identified those consistent with the measurements and clustered the plausible combinations of parameters controlling the interior.}
{We determine the plausible ranges of internal structures and the dynamical contributions to Jupiter's gravity field. Four typical interior structures are identified, characterized by their envelope and core properties. This reduces the dimensionality of Jupiter's interior to only two effective parameters. Within the reduced 2D phase space, we show that the most observationally constrained structures fall within one of the key structures, but they require a higher 1 bar temperature than the observed value.}
{We provide a robust framework for characterizing giant planet interiors with consistent wind treatment, demonstrating that for Jupiter, wind constraints strongly impact the gravity harmonics while the interior parameter distribution remains largely unchanged. Importantly, we find that Jupiter's interior can be described by two effective parameters that clearly distinguish the four characteristic structures and conclude that atmospheric measurements may not fully represent the entire envelope.}
\titlerunning{}
\authorrunning{Ziv, M., et al.}
\maketitle

\section{Introduction}

Unveiling Jupiter's internal structure is key for studying and constraining its formation and evolution with implications for other giant planets \citep{Vazan2018, Helled2022, Miguel2023, Helled2024}. NASA's Juno mission \citep{Bolton2017} provided precise measurements of Jupiter's gravity field, which are essential for constraining its interior \citep{Iess2018, Durante2020}. Additional constraints come from atmospheric measurements taken by the Galileo entry probe and Juno \citep{vonZahn1998, Seiff1998, Wong2004, Li2020}, and the cloud-level temperature derived from Voyager radio occultations \citep{Gupta2022}. 

Early Juno measurements suggested the need for a dilute core in Jupiter to match the hemispherically symmetric gravity field \citep{Wahl2017}, but this came at the expense of incompatibility with atmospheric composition measurements. To address this, an inward decrease in the heavy element mass fraction, Z (or metallicity), was proposed \citep{Debras2019}, which was shown to be unlikely from an evolution modeling perspective \citep{Howard2023c}. Other approaches to fit the atmospheric metallicity involved either lowering internal densities by increasing the temperature at 1 bar \citep{Miguel2022, Howard2023a}, or arbitrarily modifying the equation of state (EOS) for hydrogen and helium \citep{Nettelmann2021, Howard2023a}. More recent studies considered the existence of a radiative layer that could separate Jupiter's upper atmosphere from the deep Z-poor interior, allowing for atmospheric enrichment in metallicity \citep{Howard2023c, Muller2024}.

Juno also revealed hemispherical asymmetries in Jupiter's gravity field, which are attributed to deep winds that significantly affect both the symmetric and asymmetric components of the gravitational signature, adding constraints to the range of possible internal structures  \citep{Kaspi2018, Kaspi2023, Guillot2018}. These internal flows were shown to reach a depth of $\sim$$3000$ km within Jupiter \citep{Kaspi2018, Kaspi2023}. Most interior models account for the dynamical contribution to the gravity field ($\Delta J_{2n}$) by allowing it to have some typical ranges \citep{Debras2019, Miguel2022, Howard2023a}, which do not necessarily reflect the plausible solutions for each specific model and its solid-body properties. \cite{Militzer2022} takes an alternative approach, optimizing the cloud-level wind and wind decay profile for each interior model and its background density, allowing for very large $\Delta J_{2n}$ by arbitrarily varying wind decay depth with latitude.

Modeling the interior of a gas planet constrained by its gravity field is a multidimensional, nonlinear, non-unique problem that is traditionally analyzed statistically through per parameter distributions and pair-wise correlations (e.g., \citealp{Miguel2022, Howard2023a}), or by selecting the best-fitting (preferred) models (e.g., \citealp{Debras2019, Militzer2022}). Identifying typical, characteristic interior structures (i.e., combinations of interior parameters) can provide further information on the wide multidimensional solution phase space and highlight effective parameters that can simplify the representation of plausible structures, making the search for them more effective, reducing the high computational costs involved \citep{Ziv2024}.

In this study, we propose a fully consistent coupled interior and wind model where each interior model, computed using the accurate concentric Maclaurin spheroid (CMS) method \citep{Hubbard2013}, is uniquely matched to Jupiter's observed gravity field by using its own solid-body solution and a physically viable wind model. This method enables the generation of a large sample of plausible dilute core models for Jupiter, defining the permissible range for the dynamical contribution to the gravity field. Then, performing clustering analysis across the high-dimensional space of interior models simplifies them into a 2D phase space. This framework can also be adapted to explore the interiors of other giant planets.

In Sect. \ref{sect2}, we describe our modeling setup and the process used to generate a sample of plausible interior structures. Section \ref{sect3} focuses on analyzing the range of plausible interior models and their associated observables. Section \ref{sect4} presents a clustering analysis to identify key interior structures of Jupiter and shows how these structures simplify the internal model's dimensionality. We conclude in Sect. \ref{conclusion}.

\section{Methods}\label{sect2}

In this section, we build on the methodology developed by \cite{Ziv2024}, but also incorporated a wind model to obtain a sample of feasible solutions. Figure \ref{fig: Model_schematic} illustrates the interior structure model and the schematic workflow used in this study. Details of the computational steps are outlined below and in Sect. \ref{sect4.1}.

\subsection{Jupiter interior structure model}\label{sect2.1}

\begin{figure}
    \centering
    \includegraphics[width=1\linewidth]{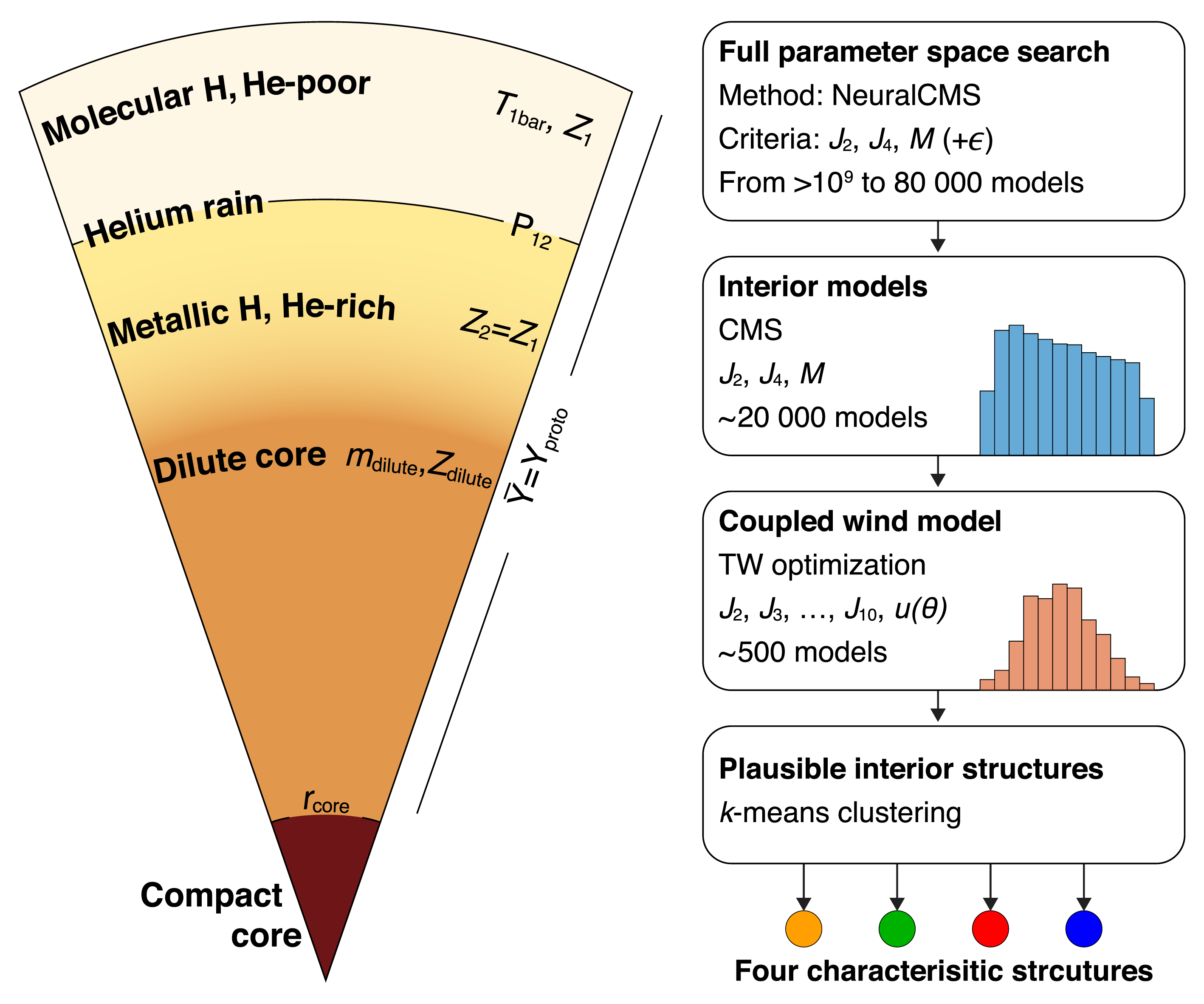}
    \caption{Schematic view of Jupiter's interior model used in this study (\textit{left}) and the exploration workflow (\textit{right}). The model's free parameters are shown. For each step of the exploration, we state which method is being used, which observables are used to constrain the solutions, and the number of plausible models resulting from this step. The prediction errors of NeuralCMS are denoted by $\epsilon$ \citep{Ziv2024}.}
    \label{fig: Model_schematic}
\end{figure}

We calculated interior models using the CMS method \citep{Hubbard2012, Hubbard2013}, based on a publicly available GitHub code\footnote{\url{https://github.com/nmovshov/CMS-planet}} \citep{Movshovitz2020}. The CMS method assembles a rotating fluid planet from $N$ constant density spheroids, constructed by their equatorial radii to compute the planetary mass, shape, moment of inertia, and gravitational moments by solving each spheroid's gravitational and rotational hydrostatic equilibrium. Moreover, radial profiles of various physical properties are computed (e.g., density, pressure, entropy, and composition). The results of CMS are sensitive to the spacing grid of spheroids \citep{Debras2018}, so we used the same equatorial grid as in \cite{Howard2023a} with $N=1041$ spheroids, in similarity to the exponential grid shown to best reproduce Jupiter analytical polytropic solution \citep{Debras2018}. The equatorial radius measured at 1 bar pressure, $R_{\rm{eq}}=71\,492\, \mathrm{km}$ \citep{Lindal1992} is unaltered in our models, so as Jupiter's rotation period of $\Omega=9.92492$ h \citep{Riddle1976}.

The gravitational moments, representing the planet's shape and mass distribution, are precisely measured by Juno \citep{Durante2020}:
\begin{equation}
    J_{n}=-\frac{1}{a^{n}M}\int r^{n}P_{n}\left(\sin \theta\right)\rho\left(r,\theta\right)d^{3}r,
\label{eq:Jn}
\end{equation}
where $n$ is the harmonic degree, $a=R_{\rm{eq}}$ is Jupiter's equatorial radius, $M$ is the planetary mass, $r$ is the radial coordinate, $\theta$ is the latitude, $P_n$ is the $n$th Legendre polynomial, and $\rho$ is the density which can be expressed as $\rho=\rho_{\rm{s}}+\rho'$, where $\rho_{\rm{s}}$ is the density from solid-body rotation (calculated by the CMS method), and $\rho'$ is the dynamical density due to the wind \citep{Kaspi2010a}. The normalized moments of inertia (NMoI$\equiv\frac{C}{MR^2_{\rm{eq}}}$, where $C$ is the moment of inertia about the rotation axis) is computed according to Eq. (5) from \cite{Hubbard2016}.

Following the setup of \cite{Miguel2022} and \cite{Howard2023a}, we modeled Jupiter's internal structure with four layers: an outer envelope, an inner envelope, a dilute core, and a possible compact core (see Fig. \ref{fig: Model_schematic}). The outer envelope is assumed to be isentropic, with its adiabat determined by the temperature at 1 bar, $T_{\rm{1bar}}$. This temperature, acting as a proxy for the deep interior entropy, was measured by the Galileo entry probe as $166.1 \pm 0.8\,\rm{K}$ and reanalyzed from Voyager radio occultations to reach $170.3 \pm 3.8\,\rm{K}$. We set the hydrogen and helium abundances to be consistent with the Galileo probe atmospheric measurements $Y_{1}/(X_{1}+Y_{1})=0.238$, where $X_1$ and $Y_1$ are the mass fractions of hydrogen and helium in the outer envelope, respectively \citep{vonZahn1998}. The envelope's metallicity, $Z_1$, was measured in the atmosphere to exceed the solar abundance \citep[see Fig. 1 in][and references therein]{Howard2023a}. Recent interior models continue to struggle with meeting this observational constraint \citep{Howard2023c}. 

The lower-than-protosolar helium abundance measured in Jupiter's atmosphere by the Galileo probe \citep{vonZahn1998}, assuming the planet's overall helium abundance is protosolar, suggests a deep enrichment of helium within the planet. A region where hydrogen and helium are immiscible could separate the He-poor outer envelope from the He-rich inner envelope, with inward helium enrichment potentially caused by "helium rain" \citep{Stevenson1977b, Mankovich2020, Howard2024}. While the phase transition pressure defining this region, $P_{12}$, remains uncertain, numerical simulations suggest it ranges between 0.8 and 3 Mbar \citep{Morales2013, Schottler2018}. We set both envelopes to have the same metallicity, $Z_1=Z_2$. The helium mass fraction in the inner envelope and dilute core is adjusted to ensure the total planetary mean helium abundance aligns with the protosolar value of $Y_{\rm{proto}}=0.278\pm 0.006$ \citep{Serenelli2010}.

The dilute core is modeled as a region with a gradual inward increase in heavy elements, following \cite{Miguel2022}:
\begin{equation}
    \Vec{Z}=Z_{2}+\frac{Z_{\mathrm{dilute}}-Z_{2}}{2}\left[1-\mathrm{erf}\left(\frac{\Vec{m}-m_{\mathrm{dilute}}}{\delta m_{\mathrm{dil}}}\right)\right],
    \label{eq:dilute}
\end{equation}
where $Z_{\mathrm{dilute}}$ is the maximum mass fraction of heavy materials in the dilute core, $m_{\mathrm{dilute}}$ denotes the extent of the dilute core in normalized mass, and $\delta m_{\mathrm{dil}}$ controls the steepness of the $Z$ gradient. We set $m_{\mathrm{dil}}=0.075$. Formation-evolution models of Jupiter suggest that the dilute core should be relatively small, extending up to 20\% of Jupiter’s mass \citep{Muller2020}. Most current models do not feature a sufficiently small dilute core, with only models from \cite{Howard2023a} matching this theoretical constraint. We explored interior models with dilute cores extending between $11\%-60\%$ of Jupiter's mass, with maximum metallicity ranging from 0.06 to 0.45. Finally, a compact core composed entirely of heavy elements is considered, defined by its normalized equatorial radius, $r_{\mathrm{core}}$, with its mass, $M_{\rm{core}}$, determined through the CMS calculation. Recent studies resulted in compact cores sized $0-8$ $M_{\oplus}$ (e.g., \citealp{Nettelmann2021, Miguel2022, Howard2023a}). We explored interior models with a compact core ranging between $0-12\%$ Jupiter's radius, equivalent to masses of $0-9.6$ $M_{\oplus}$.

The variety of EOSs for hydrogen and helium being used and uncertainties related to their interpolation, provide further uncertainty in Jupiter's plausible structures. This has been extensively studied by examining how using various EOSs affects the resulting interior models \citep{Miguel2022, Howard2023a}. In this study, we adopted the state-of-the-art pure H and He from \cite{Chabarier2019}, and accounted for their non-ideal interaction using the tables from \cite{Howard2023b}. For a given pressure and temperature, we calculate densities using the additive volume law (AVL) including the mixing effects \citep{Howard2023b, Howard2023c}:
\begin{equation}
    \frac{1}{\rho(P,T)}=\frac{X}{\rho_{{\rm H}}(P,T)}+\frac{Y}{\rho_{{\rm He}}(P,T)}+XYV_{{\rm mix}}(P,T)+\frac{Z}{\rho_{Z}(P,T)},
    \label{eq:AVL_rho}
\end{equation}
where $\rho_{\rm{H}}, \rho_{\rm{He}}, \rho_{Z}$ and $X, Y, Z$ are the pure species densities and abundances of hydrogen, helium, and heavy elements, respectively. The volume change resulting from interactions between H and He is accounted for by $V_{\rm{mix}}$. The entropy is similarly computed using the AVL:
\begin{equation}
    S(P,T)=XS_{{\rm H}}(P,T)+YS_{{\rm He}}(P,T)+XYS_{{\rm mix}}(P,T)+ZS_{Z}(P,T),
    \label{eq:AVL_S}
\end{equation}
where $S_{\rm{H}}, S_{\rm{He}}, S_{Z}$ are the pure species entropies. The entropy change due to mixing of H and He is represented by $S_{\rm{mix}}$. We used the Sesame water EOS \citep{Lyon1992} to account for the heavy elements. To compute the densities $\rho$ (in $\rm{g\,cm^{-3}}$) as a function of pressure $P$ (in Mbar) in the compact core we used the "rock" analytical formula from \cite{Hubbard1989}:
\begin{equation}
    P=\rho^{4.406}\exp(-6.579-0.176\rho+0.00202\rho^{2}).
    \label{eq:EOS_core}
\end{equation}

\subsection{Exploration of plausible interior models with NeuralCMS}\label{sect2.2}

{\renewcommand{\arraystretch}{1.2}
\begin{table}
    \caption{Observables used to constrain the interior models.}
    \label{tab:Criteria}
    \centering
    \begin{tabular}{l c}
    \hline\hline
     Criteria & Constraints \\
    \hline
        Interior & $\lvert J_{2}^{{\rm CMS}}-J_{2}^{{\rm Juno}}\rvert\leq2\times10^{-6}$ \\
        
         & $\lvert J_{4}^{{\rm CMS}}-J_{4}^{{\rm Juno}}\rvert\leq10^{-6}$ \\ 
         & $\lvert M^{{\rm CMS}}-M^{{\rm Juno}}\rvert\leq\Delta M=0.0005\times10^{27}\,{\rm kg}$ \\
    \hline
        Wind & $\lvert(\Delta J_{2n}^{{\rm sol}}+J_{2n}^{{\rm CMS}})-J_{2n}^{{\rm Juno}}\rvert\leq3\sigma_{2n}^{{\rm Juno}}\,(n=1,\ldots,5)$\\
        & $\lvert\Delta J_{n}^{{\rm sol}}-J_{n}^{{\rm Juno}}\rvert\leq3\sigma_{n}^{{\rm Juno}}\,(n=3,5,7,9)$\\
        & $\lvert u^{{\rm sol}}(\theta)-u^{{\rm obs}}(\theta)\rvert\leq20\,{\rm m\,s^{-1}}$\\
    \hline
        Obs. &  $T_{{\rm 1bar}}\leq178\,{\rm K}$ \\
         & $Z_{1}\geq0.015\thickapprox Z_{{\rm solar}}$ \\
         & $P_{12}\leq3\,{\rm Mbar}$ \\
    \hline
    \end{tabular}
    \tablefoot{The CMS solutions are the static gravity harmonics. The Juno-measured gravity harmonics ($J_{n}^{{\rm Juno}}$) and their corresponding $3\sigma$ uncertainties ($3\sigma_{n}^{{\rm Juno}}$) are taken from \cite{Durante2020}. The Juno-derived mass is $M^{\rm{Juno}}=1.8983\times10^{27}\, {\rm kg}$, and $\Delta M$ is the mass uncertainty due to the available range of the gravitational constant $G$ \citep[see][]{Ziv2024}. The observed cloud-level wind $u^{\rm{obs}}(\theta)$ are from \cite{Tollefson2017}. We consider $Z_{{\rm solar}}$ as the protosolar value from \cite{Lodders2009}.}
\end{table}
}

We reduced Jupiter's interior structure to seven controlling variable parameters discussed in Sect. \ref{sect2.1}: $T_{\rm{1bar}}$, $Z_{1}$, $P_{12}$, $Y_{\rm{proto}}$, $m_{\mathrm{dilute}}$, $Z_{\mathrm{dilute}}$, and $r_{\mathrm{core}}$. To explore the phase space of these parameters and find parameter combinations consistent with Juno gravity data, we used NeuralCMS\footnote{\url{https://github.com/zivmaaya/NeuralCMS}}, a deep learning approach regressing the CMS model \citep{Ziv2024}. NeuralCMS is a deep neural network model, predicting the even gravitational moments $J_2$ to $J_8$ and the mass given a combination of the seven interior parameters mentioned above. It is trained on a large sample of wide-ranged CMS-computed interior models of Jupiter allowing the performance of an extensive grid search for plausible interior models with a reduced computational runtime by a factor of $10^5$. NeuralCMS was applied to eliminate models inconsistent with Juno and reduce the parameter range. Due to the small measurement uncertainty of $J_2$ and $J_4$, actual CMS calculations were performed to exclude models falsely predicted to be plausible. 

The grid search procedure follows the approach described in \cite{Ziv2024}, where all parameter combinations are explored using a uniformly spaced grid for each parameter with $m$ grid points, resulting in $m^{7}$ interior models per grid search iteration, without prior assumptions on the model parameters. Models consistent with Juno's observations are selected based on the interior criteria in Table \ref{tab:Criteria}, added with a prediction error from NeuralCMS, $\epsilon$, to the allowable deviation from Juno's measurements. The interior criteria allow interior models to deviate significantly from the Juno $3\sigma$ uncertainty on $J_{2}$ and $J_4$, which is $\sim$$2\times10^{-9}$ to investigate the permissible range of interior models allowed by a self-consistent wind model. 

In this study, we build upon the two grid search iterations conducted by \cite{Ziv2024}. The initial iteration considers the maximum prediction errors on the dataset used to train NeuralCMS, to roughly reduce the range of the interior parameters. The second grid search uses a denser grid, the $3\sigma$ prediction errors on the training dataset, $\epsilon_{3\sigma}$, and the reduced parameter range, followed by CMS calculations to retrieve a sample of plausible interior models. To obtain a larger sample, a third grid search iteration was performed with the further narrowed parameter range from the second iteration, $\epsilon_{3\sigma}$, and an even denser grid. The third and final iteration with NeuralCMS yielded approximately $80\,000$ plausible interior models within $\epsilon_{3\sigma}$, which were recalculated with CMS to retrieve a sample of $\sim$$20\,000$ parameter combinations that satisfy the interior criteria (Table \ref{tab:Criteria}). The three grid search stages are summarized in Table \ref{tab:Grid_search}. We confirmed that NeuralCMS does not add uncertainty to the distribution of interior parameters by verifying that no plausible models were excluded by it. This was done by recalculating (with the CMS method) the third grid search iteration with a larger prediction error margin, close the maximum errors on the validation dataset for $J_2$, $J_4$, and $M$.

\subsection{Coupled thermal wind model with modified surface wind}\label{sect2.3}

\begin{figure}
    \centering
    \includegraphics[width=1\linewidth]{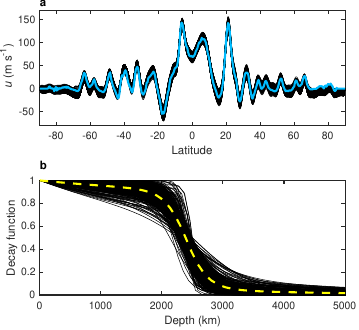}
    \caption{Optimized wind solutions for all models accepted by the wind criteria (Table \ref{tab:Criteria}). \textit{Panel a}: latitudinal cloud-level wind profiles. \textit{Panel b}: the radial decay function with depth. The blue profile represents the observed cloud-level wind \citep{Tollefson2017}, and the dashed yellow profile indicates the mean of all decay profiles.}
    \label{fig: Wind_solutions}
\end{figure}

The observed cloud-level wind and gravity data impose constraints on the flow dynamics and, consequently, the dynamical contribution to Jupiter's gravity field. In fast-rotating giant planets like Jupiter, large-scale flows are predominantly governed by geostrophic balance \citep{Pedlosky1987, Kaspi2009}. Given the largely zonally symmetric nature of Jupiter's flow, the leading order dynamical balance is between the flow gradient in the direction parallel to the axis of rotation and the latitudinal density perturbations \citep{Galanti2017a}, given by a thermal wind (TW) balance:

\begin{equation}
    2\Omega r\frac{\partial}{\partial z}\left(\rho_{\rm{s}}u\right)=\textsl{g}_{\rm{s}}\frac{\partial\rho'}{\partial\theta},
    \label{eq:TW}
\end{equation}
where $u$ is the wind, $\rho_{\rm{s}}\left(r\right)$ and $\textsl{g}_{\rm{s}}\left(r\right)$ are the static (solid-body) density and gravity fields, computed with CMS, $z$ is the direction parallel to the axis of rotation, and $\rho'\left(r,\theta\right)$ is the anomalous density field. For a detailed derivation of this form of the TW balance, see \cite{Kaspi2018}. The wind-induced gravitational moments are then calculated based on the anomalous density field
\begin{equation}
    \Delta J_{n}=-\frac{2\pi}{MR_{{\rm eq}}^{n}}\intop_{-\pi/2}^{\pi/2}\cos\theta d\theta\intop_{0}^{R_{{\rm eq}}}r^{n+2}P_{n}\left(\sin\theta\right)\rho'\left(r,\theta\right)dr,
    \label{eq:Delta_Js}
\end{equation}
where $n=2,3,\ldots,N$. This balance was used to determine the depth of Jupiter's winds using Juno's gravity data \citep{Kaspi2018}, and similarly for Saturn using Cassini's gravity data \citep{Galanti2019a}.

In this study, we followed the methodology of \cite{Galanti2019a}, where the cloud-level wind profile and its decay are adjusted to best fit the Juno-measured even and odd gravity harmonics, $J_{2}$ to $J_{10}$. We accounted for uncertainties in the observed cloud-level wind \citep{Tollefson2017}, by selecting plausible models that are within $20\,\rm{m}\,\rm{s}^{-1}$ of the observed wind (see Fig. \ref{fig: Wind_solutions}a). Firstly, we decompose the observed wind into the first $N=99$ Legendre polynomials $u^{{\rm obs}}\left(\theta\right)=\sum_{i=0}^{N}A_{i}^{{\rm obs}}P_{i}\left(\sin\theta\right)$, where the coefficients $A_{i}^{\rm{obs}}$ set the latitudinal wind profile. During optimization, these coefficients are allowed to vary, resulting in a modified cloud-level wind, $u^{{\rm sol}}\left(\theta\right)=\sum_{i=0}^{N}A_{i}^{{\rm sol}}P_{i}\left(\sin\theta\right)$. Next, the modified wind $u^{\rm{sol}}\left(\theta\right)$ is projected parallel to the axis of rotation and then decayed radially using a continuous, monotonic decay function that is a linear combination of exponential and normalized hyperbolic tangent functions (see Fig. \ref{fig: Wind_solutions}b). The wind depth remains constant across all latitudes. Projecting parallel to the rotation axis and applying radial decay have been demonstrated to provide the most probable flow structure for explaining the gravity measurements \citep{Kaspi2023}. We note that \cite{Militzer2022} also uses this procedure, in addition to another degree of freedom that allows for latitude-dependent wind decay.

The CMS calculation provides radial profiles of each interior model's equipotential surfaces, density, temperature, and composition. These profiles serve as the background (static) state for each interior model, which is then used in Eq. \ref{eq:TW}, to couple the interior model with a wind model. This approach allows for the consistent selection of interior models that match with the observed cloud-level wind within its uncertainties and the Juno-measured gravitational moments within Juno's $3\sigma$ uncertainty \citep{Durante2020}. We applied the wind model to the $\sim$$20\,000$ models accepted by the interior criteria discussed in Sect. \ref{sect2.2} (shown as blue histograms in Fig. \ref{fig: Distributions}) and identified 491 plausible interior structures that satisfy both the interior and the wind criteria (see Table \ref{tab:Criteria}). These wind-constrained models are shown in red histograms in Fig. \ref{fig: Distributions}.

\section{Results I: Plausible range of interior structures}\label{sect3}

\begin{table*}
    \caption{Results for the dynamical contribution to the even gravity harmonics ($\Delta J_{2n}=J_{2n}^{\rm{Juno}}-J_{2n}^{\rm{static}}$) compared with previous studies.}
    \label{tab:Delta_Js2}
    \centering
    \begin{tabular}{l r r r r}
    \hline\hline
      & \multicolumn{1}{c}{This work $1\sigma$} & \multicolumn{1}{c}{\cite{Miguel2022}} & \multicolumn{1}{c}{\cite{Kaspi2020}} & \multicolumn{1}{c}{\cite{Galanti2021}} \\
    \hline
        \(\Delta J_{2}\times10^{6}\)\ & $0.530\pm 0.214$ & $1.039\pm 0.354$ & $0.558$ & $0.396$\\
        \(\Delta J_{4}\times10^{6}\)\ & $-0.090\pm 0.114$ & $-0.076\pm 0.083$ & $-0.048$ & $-0.031$\\
        \(\Delta J_{6}\times10^{6}\)\ & $-0.111\pm 0.041$ & $0.016\pm 0.076$ & $0.010$ & $-0.003$\\
        \(\Delta J_{8}\times10^{6}\)\ & $0.046\pm 0.007$ & $0.053\pm 0.062$ & $0.035$ & $0.040$\\
        \(\Delta J_{10}\times10^{6}\)\ & $-0.041\pm 0.003$ & $-0.080\pm 0.042$ & $-0.030$ & $-0.031$\\
    \hline
    \end{tabular}
\end{table*}

Figure \ref{fig: Distributions} presents the distributions of interior model parameters and observables for the plausible interior structures that satisfy the interior criteria (blue histograms) and the subset that additionally meets the wind criteria (red histograms). In this section, we examine the differences between these two sets of models and analyze the resulting ranges of parameters and observables. Figure \ref{fig: Corner_plot} shows the pair-wise relations (corner plot) of all 491 plausible interior models.

\subsection{Distribution of the observables}\label{sect3.1}

\begin{figure*}
    \centering
    \includegraphics[width=1\linewidth]{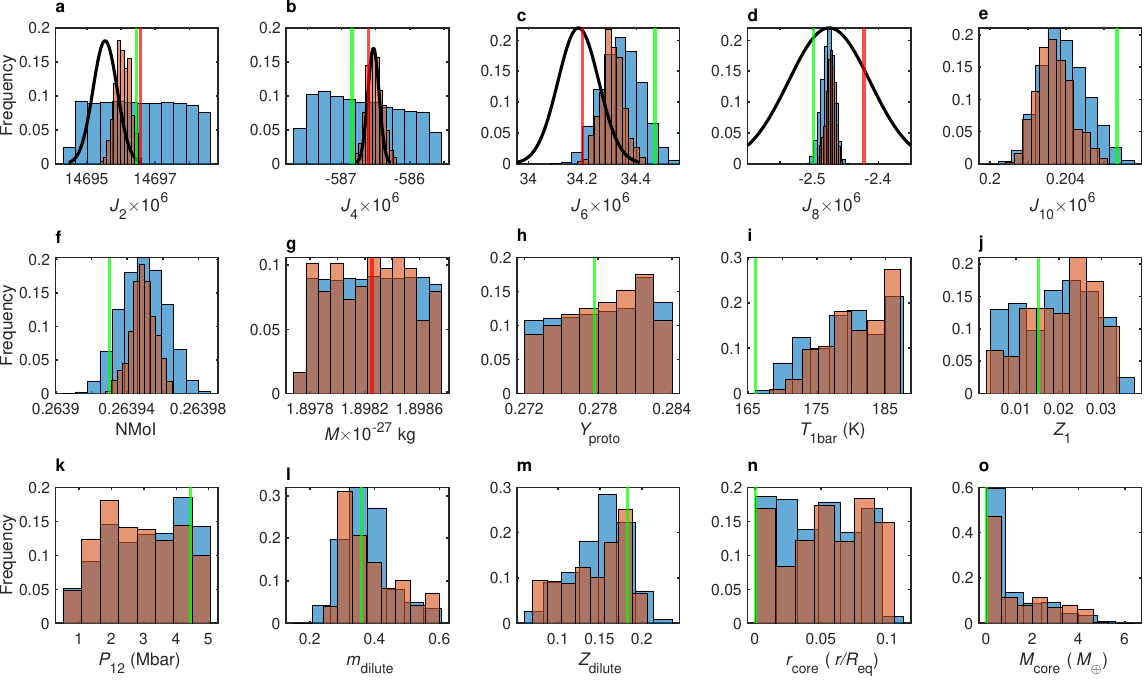}
    \caption{Distribution of observables (\textit{a-g}) and interior structure parameters (\textit{h-n}) for plausible interior models. The gravity harmonics (\textit{a-e}) are the static components, which are fitted to Juno's measurements using the wind model. \textit{Panel o} shows the compact core mass determined by $r_{\rm core}$. Blue histograms correspond to models that satisfy the interior criteria ($19\,982$ models), while red histograms represent those that also meet the wind criteria (491 models). The red vertical line marks Juno's $J_{2n}$ measurements \citep{Durante2020} and derived mass \citep{Ziv2024}. Black Gaussians represent the allowed range for the static gravity harmonics accounting for differential rotation ($J_{2n}^{\rm static}=J_{2n}^{\rm Juno}-\Delta J_{2n}^{\rm dynamical}$) from \cite{Miguel2022}. For $J_{10}$ (\textit{e}), the Juno measurement lies outside the range shown, but the black Gaussian covers the displayed values. The preferred model from \cite{Militzer2022} is shown in green lines (we show the inner edge of the He rain region compared to $P_{12}$). The gravity harmonics (\textit{a-e}) distance from the red lines represents $\Delta J_{2n}$. The histogram color corresponds to Fig. \ref{fig: Model_schematic}.}
    \label{fig: Distributions}
\end{figure*}

We find that the lower gravity harmonics $J_2$ and $J_4$ (Fig. \ref{fig: Distributions}a,b) are strongly constrained by applying the wind model, indicating that the explored range was sufficiently broad. The wind constraint is different from that used by \cite{Miguel2022} (black lines in Fig. \ref{fig: Distributions}), where they examined the plausible dynamical contribution to the even gravity harmonics through random sampling of $\Delta J_{2n}$, using a constant background state for the TW optimization described in Sect. \ref{sect2.3}. They selected plausible solutions based on similar wind criteria to those used in this study (Table \ref{tab:Criteria}). We note that the results from \cite{Miguel2022} presented here represent only the wind-allowed range for the static $J_{2n}$, not the distribution of plausible interior models. The comparison here is therefore intended for these allowable ranges, as also shown in Table \ref{tab:Delta_Js2}. Comparing our results to theirs, we observe a different range for $\Delta J_{2}$ and a comparable distribution for $\Delta J_{4}$ (see Fig. \ref{fig: Distributions}a,b and Table \ref{tab:Delta_Js2}). The higher-order even gravity harmonics exhibit a narrower range due to the applied interior criteria, meaning that their values are largely determined by the interior structure. For $\Delta J_{6}$, the observable which is more challenging to fit \citep{Debras2019, Militzer2022}, we find a different range than \cite{Miguel2022} with the largest allowed deviation from Juno to be $\Delta J_{6}\times10^{6}=-0.22$ (for an interior model with $T_{\rm{1bar}}=171\,\rm{K}$ and $Z_{1}=0.009$). For $J_{8}$ and $J_{10}$ we observe much tighter distributions compared to \cite{Miguel2022}. Table \ref{tab:Delta_Js2} also compares our findings with two previous studies that determine the permissible range of dynamical contribution to the gravity harmonics matching the gravity field observations \citep{Kaspi2020}, and also magnetic field constraints \citep{Galanti2021}. Our results are more consistent with these studies than with \cite{Miguel2022}. For $J_{6}$, our deviations from the Juno measurements are larger than in the previous studies, consistent with the finding that the best-fit wind solutions favor smaller values for $\Delta J_{6}$ \citep{Kaspi2023}.

During our exploration, we allowed the planetary mass to vary within the uncertainty stemming from various values reported for the gravitational constant $G$ \citep{Tiesinga2021, Ziv2024}. This is similar methodologically to allowing variations in the equatorial radius while keeping the mass constant, as done by  \cite{Howard2023a} and \cite{Miguel2022}. Figure \ref{fig: Distributions}g shows that the mass was not constrained by the wind model. Similarly to the gravity harmonics higher-order than $J_{4}$, the NMoI is primarily determined by the interior structure, and the wind model further constrains it. We find values of $\rm{NMoI=0.26395\pm0.00002}$ ($3\sigma$ uncertainty) for the sample of 491 plausible models. This result is consistent with \cite{Militzer2023a}, who derived values of $0.26393-0.26398$ with abstract models that match Juno's $J_{2}-J_{6}$, and the range for the preferred model from \cite{Militzer2022}, with $0.26393\pm0.00001$ (shown in green line in Fig. \ref{fig: Distributions}), which is on the edge of our allowed range. \cite{Neuenschwande2021} presented empirical structure models of Jupiter matching the measured $J_{2}$, $J_{4}$, and equatorial radius, but with a lower derived NMoI of $0.263408-0.263874$.

The preferred model from \cite{Militzer2022}, which matches Jupiter's gravity field by allowing the wind depth to vary with latitude, is represented by the green lines in Fig. \ref{fig: Distributions}. Their study indicated that if the wind depth is kept constant across latitudes, the optimized cloud-level wind deviates by $50\,\rm{m}\,\rm{s}^{-1}$ from the observed values, much more then the observed uncertainties \citep{Tollefson2017}, and well outside our wind criteria. Our modeling setup is different than their preferred model by the exact helium rain and the dilute core regions implementation, the EOS used for H-He mixture, where they use the tables from \citep{Militzer2014}, but most importantly the wind approach. We show in Fig. \ref{fig: Distributions}a-f that their model is either on the edge of our allowed range for the gravitational moments and NMoI, or falls outside it. We find that while their model meets the interior criteria for the even gravity harmonics, it is excluded as a plausible structure after applying the coupled wind model. Fitting $J_{6}$ is challenging with most available EOSs, as demonstrated by \cite{Howard2023a}, who showed that achieving a good fit requires EOSs associated with high $T_{\rm{1bar}}$ values, above 180 K, much higher than the Galileo probe measurement of 166.1 K imposed by \cite{Militzer2022}.

\subsection{Distribution of the interior physical parameters}\label{sect3.2}

Next, we present the range of parameters defining Jupiter’s interior, along with the resulting mass and distribution of heavy elements within the planet. First, it is important to note that NeuralCMS significantly narrows the ranges of parameters such as $T_{\rm{1bar}}$, $Z_{1}$, $m_{\rm{dilute}}$, $Z_{\rm{dilute}}$, and $r_{\rm{core}}$ \citep{Ziv2024}. Figure \ref{fig: Distributions}h-o shows that the distribution of the interior parameters remains largely unchanged when wind constraints are applied. We note that the peak values in the histograms do not necessarily correspond to a plausible combination of parameters. Our models indicate a preference for high $T_{\rm{1bar}}$, and small compact core mass ($M_{\rm{core}}$) ranging from 0 to 5.3 $M_{\oplus}$, which is consistent with findings from previous studies \citep{Nettelmann2021, Miguel2022, Howard2023a}. Additionally, $T_{\rm{1bar}}$ and $Z_{1}$ exhibit strong positive correlation \citep{Ziv2024}, resulting in a higher number of models with increased envelope metallicity, reaching up to $Z_{1}=0.0342$ (2.3 times the solar value), associated with $T_{\rm{1bar}}=187\,\rm{K}$. 

The characteristics of the dilute core in our models are consistent with those calculated using the same EOS from \cite{Howard2023a}, though with a different setup. The plausible range of $m_{\rm{dilute}}$ is found to be between $0.25-0.6$ and between $0.072-0.204$ for $Z_{\rm{dilute}}$. We report values for the masses of heavy elements in the different regions in the interior (also shown in Fig. \ref{fig: Corner_plot}) defined the same as in \cite{Howard2023a}. The mass of heavy elements in both envelopes, corresponding to the envelope metallicity and not the enrichment in the dilute core $M_{Z,\rm{env}}$ ranges from 1.6 to 10.9 $M_{\oplus}$ and is primarily determined by the value of $Z_{1}$. The excess mass of heavy elements in the dilute core region $M_{Z,\rm{dil}}$ ranges from 7.6 to 19.6 $M_{\oplus}$. The total mass of heavy elements in the planet $M_{Z,\rm{total}}$ varies between 17.9 and 25.8 $M_{\oplus}$. This means that our models have most of their heavy elements in the dilute core region instead of in the envelope, in agreement with \cite{Howard2023a}.

\section{Results II: Characteristic interior structures}\label{sect4}

The range of plausible interior structures for Jupiter, shaped by the complex interplay of properties like composition, compositional gradients, and thermal structure, makes it challenging to identify typical structures. However, determining such representative structures can simplify our characterization of Jupiter’s interior, provide valuable multidimensional insights, and enable meaningful comparisons with other giant planets.

\subsection{Clustering analysis based on the interior parameters}\label{sect4.1}

\begin{figure}
    \centering
    \includegraphics[width=1\linewidth]{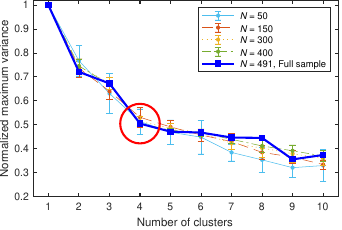}
    \caption{Normalized maximum variance for increasing number of clusters tested. Shown is the analysis for the full sample of plausible interior structures (thick blue), and the mean and standard deviation of the analysis for 10 randomly selected sub-samples with different sizes. The red circle marks our selection of four clusters for this analysis.}
    \label{fig: Elbow_plot}
\end{figure}

The variability among plausible interior structures can be explored and simplified through clustering analysis, which helps identify characteristic interior structures of Jupiter by classifying common combinations of parameters. In this study, we used the $k$-means method which was successfully used to characterize the behavior of Jupiter's magnetosphere \citep{Collier2020}. We used the sample of 491 plausible interior models with their seven defining parameters (Fig. \ref{fig: Model_schematic}) for this analysis. To avoid biases, we standardized the parameters by subtracting the mean and dividing by the standard deviation.

The $k$-means algorithm starts by randomly selecting $k$ cluster centroids and assigning each of the 491 data points to the nearest centroid based on Euclidean distance. The centroids are then updated as the mean of their assigned points. This process repeats until centroids no longer change. To ensure robust results, the algorithm runs multiple times with different initializations, selecting the setup with the lowest sum of squared distances. The analysis was performed using the Matlab "kmeans" function\footnote{\url{https://www.mathworks.com/help/stats/kmeans.html}}. The optimal number of clusters was determined by an "elbow analysis", which shows how variance decreases with an increasing number of clusters (Fig. \ref{fig: Elbow_plot}). In our case, variance drops sharply up to four clusters, with only marginal improvement beyond that. Adding more than four clusters yields little additional simplicity despite a minor variance reduction at nine clusters. 

To demonstrate that the sample size of 491 is sufficient for statistical analysis, we performed the same analysis on sub-samples of varying sizes, as shown in Fig. \ref{fig: Elbow_plot}. Our results indicate that for sub-samples with fewer than $N=150$ data points, the variability in the elbow analysis is too high to statistically validate the quality of the clustering. Additionally, for these smaller sub-samples, there is a significant reduction in variance when using more than four clusters, which is not observed in larger sub-samples. In contrast, for the larger sub-samples, the behavior of the elbow plot closely resembles that of the full sample, supporting the robustness of our clustering analysis.

\begin{figure}
    \centering
    \includegraphics[width=1\linewidth]{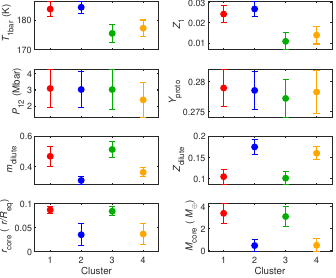}
    \caption{Interior parameters' means (points) and standard deviations (error bars) within the four clusters, shown in different colors. Clusters 1 and 2 (red and blue) show high values of $T_{\rm 1bar}$ and $Z_{1}$, while clusters 3 and 4 (green and yellow) show lower values of these parameters. Clusters 1 and 3 feature high values of $m_{\rm dilute}$ and $r_{\rm core}$, and low values of $Z_{\rm dilute}$, whereas clusters 2 and 4 display the opposite trend. We note that the compact core mass, $M_{\rm{core}}$, was not used in the clustering analysis.}
    \label{fig: Clustering_means}
\end{figure}

Figure \ref{fig: Clustering_means} shows the variability of the interior parameters across the four clusters. The parameters controlling the envelope ($T_{\rm{1bar}}$ and $Z_{1}$) are distinguished between clusters, where two clusters are characterized by high values of these parameters and the other two clusters have low values. Conversely, The transition pressure setting the helium rain region $P_{12}$, and the mean planetary helium mass fraction set by $Y_{\rm{proto}}$ are not classified within the four clusters. Similar to the envelope parameters, the parameters controlling the planet's core ($m_{\rm dilute}$, $Z_{\rm dilute}$, and $r_{\rm core}$) are also separated between clusters, but with two different clusters characterized by high (low) values for $m_{\rm dilute}$ and $r_{\rm core}$ ($Z_{\rm dilute}$), and vice versa for the other two clusters. These relationships between the core-controlling parameters are consistent with previous studies \citep{Miguel2022, Howard2023a}. Additionally, we find that the smallest plausible dilute cores are associated with a relatively hot and heavy envelope. Overall, the clusters represent four unique combinations of two envelope configurations and two core configurations. Again, we note that the parameter combination represented by the mean values of each cluster is not necessarily a plausible combination.

\subsection{Reducing the dimensionality to two effective parameters}\label{sect4.2}

\begin{figure}
    \centering
    \includegraphics[width=1\linewidth]{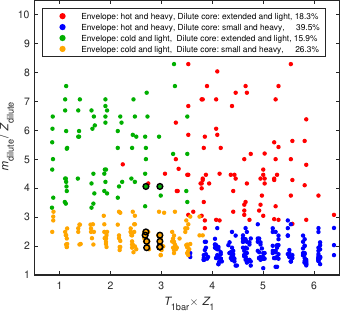}
    \caption{All 491 identified plausible interior models presented in the phase space defined by the product of the two envelope-controlling parameters and the ratio of the two dilute core-controlling parameters. Colors represent different clusters, consistent with Fig. \ref{fig: Clustering_means}. Black circles show the most observationally constrained interior models (see the observational criteria in Table \ref{tab:Criteria}). The legend displays the fraction of models assigned to each cluster. "Heavy" and "light" refer to metallicity.}
    \label{fig: Clustering_phase_space}
\end{figure}

\begin{figure*}
    \sidecaption
    \includegraphics[width=12cm]{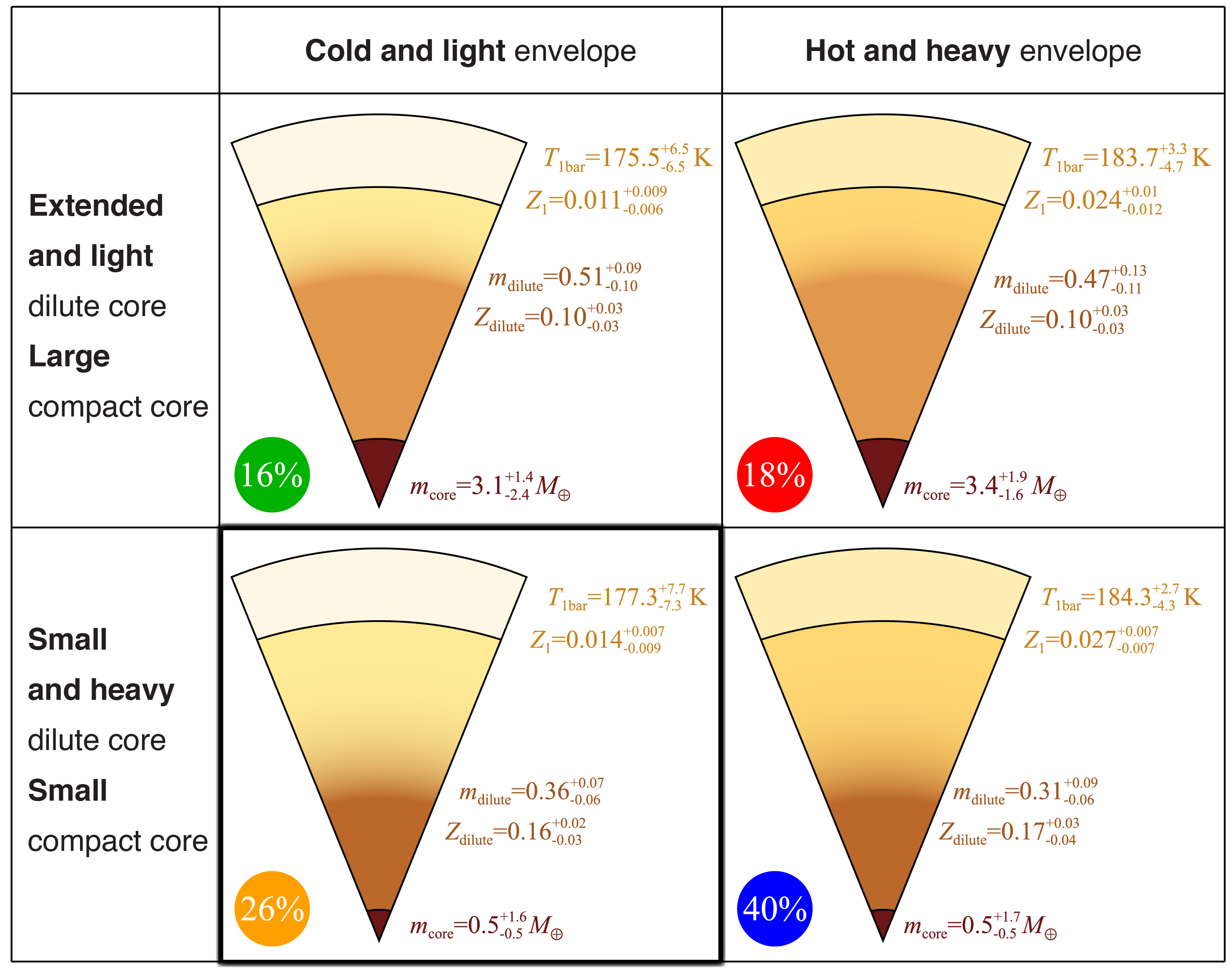}
    \caption{Schematic diagram of the four characteristic interior structures of Jupiter identified in this study. The interior parameters mean values and full range within each cluster are presented. The fraction of plausible interior models within each cluster is shown in the same color coding as in Figs. \ref{fig: Clustering_means} and \ref{fig: Clustering_phase_space}. The structure with the most observationally constrained interior models is shown with a thicker frame. The diagrams are not to scale.} 
    \label{fig: Clusters_schematic}
\end{figure*}

As discussed above, the resulting clusters are distinguished by their envelope and core states, indicating that interior structures can be effectively represented in a 2D phase space with two key parameters representing these states. Figure \ref{fig: Clustering_phase_space} presents all 491 plausible interior structures, color-coded by their assigned cluster, according to two effective parameters: the product of temperature at 1 bar and envelope metallicity (which distinguishes between hot and heavy versus cold and light envelopes), and the ratio of dilute core extent to maximum dilute core metallicity (which differentiates between extended and light versus small and heavy dilute cores). The latter parameter also correlates with the compact core radius and mass (see Fig. \ref{fig: Clustering_means}), where an extended dilute core is associated with a larger (heavier) compact core. The descriptors used (i.e., cold versus hot, small versus extended, light versus heavy) are relative to the derived plausible range (Fig. \ref{fig: Distributions}), with "light" and "heavy" referring to metallicity (molecular weight). Figure \ref{fig: Clustering_phase_space} also presents the fraction of interior models within each cluster, showing a higher density of models with a hot and heavy envelope and a small and heavy dilute core. A schematic of the four characteristic Jupiter interior structures, including the full parameter range of each cluster, is provided in Fig. \ref{fig: Clusters_schematic}. Importantly, we show that Jupiter's interior structure can be effectively explained using a 2D phase space, clearly representing the four distinguishable characteristic structures. Moreover, the overlap at the edges of clusters suggests they represent characteristic structures rather than distinct end members.

Additional observational and theoretical constraints can be applied to the sample of plausible interior structures to refine the selection. Models consistent with atmospheric measurements and theoretically bound for the helium rain region, meeting the observational criteria listed in Table \ref{tab:Criteria}, are marked with black circles in Fig. \ref{fig: Clustering_phase_space}. These criteria include a temperature at 1 bar lower than 178 K (close to the upper bound from reanalysis of the Voyager radio occultations derived by \cite{Gupta2022}), higher than solar atmospheric (envelope) metallicity, and a helium rain transition pressure lower than 3 Mbar \citep{Morales2013}. Most highlighted structures fall within a cluster characterized by a cold and light envelope and a small, heavy dilute core, suggesting that Jupiter’s interior structure might be confined to this specific characteristic structure or at least a narrow region in the phase space (Fig. \ref{fig: Clustering_phase_space}). This structure is also highlighted with a thick frame in Fig. \ref{fig: Clusters_schematic}. Under these constraints, the minimal $T_{\rm{1bar}}$ required to accommodate at least a solar abundance of heavy elements in the envelope exceeds 175 K, which is higher than the upper limit of 174.1 K derived from Voyager radio occultations \citep{Gupta2022}, suggesting that the deep interior entropy might not be represented by the measured $T_{\rm{1bar}}$.

\section{Conclusion}\label{conclusion}

In this study, we explored a wide range of Jupiter’s structural models with self-consistent interior and wind models constrained by Juno's gravity field measurements, observed cloud-level wind, and atmospheric observations from both Juno and the Galileo entry probe. These models feature four layers: an outer envelope defined by the temperature at 1 bar and atmospheric metallicity, an inner envelope with the same metallicity but enriched in helium due to helium rain, a dilute core characterized by compositional gradients, and a possible compact core.

We present a robust framework for characterizing Jupiter’s interior using machine learning techniques, which can also be applied to study the interiors of other giant fluid (exo)planets. Our approach uses NeuralCMS, a deep neural network model, to efficiently reduce the range of plausible interior structures and explore the full parameter space. Subsequently, we compute interior models using the accurate CMS method, integrated with a wind model that has been demonstrated to best explain the observed asymmetric gravity field \citep{Kaspi2023}. We used a state-of-the-art EOS for hydrogen and helium; however, using alternative available EOS could yield different plausible ranges and distributions of interior models, potentially impacting the clustering analysis. This methodology ensures a consistent selection of plausible models.

The wind constraints result in a different range of plausible dynamical contributions to the gravity field ($\Delta J_{2n}$) compared to the values found by \cite{Miguel2022}, who used random sampling of $\Delta J_{2n}$ combinations to account for the wind in their interior models, as well as by \cite{Howard2023a}. These differences may arise from variability between the background density profiles used to derive wind-induced gravity harmonics and differences in the EOS. We show that the even gravity harmonics $J_{6}$ to $J_{10}$ are predominantly determined by the interior structure, thus suggesting that our derived range more accurately reflects the wind effects relevant to interior modeling.

The application of wind constraints has further implications: while the distribution of parameters controlling the interior does not change significantly by applying the wind model, the range of plausible interior parameters is more narrowly defined. This is particularly important for addressing physical features that recent interior models have struggled to reconcile with observations and theory, such as the temperature at 1 bar, atmospheric metallicity, and the extent of the dilute core.

Clustering analysis reveals combinations of seven interior parameters with similar characteristics, offering deeper insights than traditional pair-wise relations or preferred models. We identify four characteristic interior structures of Jupiter that differ in their envelope and planetary core configurations. Specifically, the envelope is categorized as either hot and heavy or cold and light, and the core configuration varies between a small, heavy dilute core with a small compact core or the opposite (relative to the derived plausible interior range). These four different configurations are expected from a mass balance perspective.

The classification process enabled us to simplify Jupiter's interior characteristics into two effective parameters: one representing the envelope state and the other representing the core state. This reduced the dimensionality of the analysis from a 7D space to a 2D phase space, clearly highlighting the four characteristic structures. Within the reduced 2D phase space, we highlight the models most consistent with the combination of atmospheric measurements ($T_{\rm{1bar}}$ and $Z_{1}$) and show that they mostly fall within a specific characteristic cluster, thus providing a further important reduction towards one key internal model. These models are characterized by a small and heavy dilute core with mean values of $m_{\rm dilute}=0.36$ and $Z_{\rm dilute}=0.16$. Our refined results on the extent of the dilute core are consistent with previous studies by \cite{Debras2019} and \cite{Militzer2022}, but diverge from the formation-evolution models of \cite{Muller2020}, which suggest a less extended dilute core. However, even these constrained models face challenges in matching with both supersolar atmospheric metallicity and the observationally consistent 1 bar temperature, indicating that these measurements might not fully represent Jupiter’s entire envelope within this modeling setup.

\begin{acknowledgements}
We thank the referee for valuable comments and suggestions, which have improved the quality of this manuscript. We thank the Juno Interior Working Group for useful discussions and comments, and Moria Abu for her graphic assistance with the schematic figures. We acknowledge support from the Israeli Space Agency and the Helen Kimmel Center for Planetary Science at the Weizmann Institute.
\end{acknowledgements}

\bibliographystyle{aa}
\bibliography{manuscript}

\begin{appendix}

\onecolumn
\section{Grid search iterations with NeuralCMS}\label{apendix:CMS validation}

\begin{table*}[h]
    \centering
    \caption{NeuralCMS grid search stages details.}
    \label{tab:Grid_search}
    \begin{tabular}{l c c c}
    \hline\hline
      & \multicolumn{1}{c}{Iteration I} & \multicolumn{1}{c}{Iteration II} & \multicolumn{1}{c}{Iteration III}\\
    \hline
        \(T_{\rm{1bar}}\,(\rm{K})\)  & $159-185$ & $163.5-185$ & $163-187$\\
        \(Z_1\) & $0.005-0.06$ & $0.005-0.04 $ & $0.005-0.04$ \\
        \(P_{12}\,(\rm{Mbar}\)) & $0.8-5$ & $0.8-5$ & $0.8-5$ \\
        \(Y_{\rm{proto}}\)   & $0.272-0.284$ & $0.272-0.284$ & $0.272-0.284$ \\
        \(m_{\rm dilute}\) & $0.11-0.6$ & $0.14-0.6$ & $0.16-0.6$ \\
        \(Z_{\rm dilute}\) & $0.06-0.45$ & $0.06-0.28$ & $0.065-0.24$ \\
        \(r_{\rm{core}}\,(r/R_{\rm{eq}})\) & $0-0.12$ & $0-0.1125$ & $0-0.11$ \\
    \hline
        $m$ & 17 & 20 & 25 \\
        Prediction error & $\epsilon_{\rm{max}}$ & $\epsilon_{3\sigma}$ & $\epsilon_{3\sigma}$ \\
    \hline
    \end{tabular}
    \tablefoot{The parameter range is gridded with $m$ grid points per parameter, and a prediction error is added to the allowed deviation from Juno according to the interior criteria presented in Table \ref{tab:Criteria}. Plausible models in iterations II and III are recalculated with CMS.}
\end{table*}

\twocolumn
\onecolumn
\section{Corner plot of plausible interior models}\label{apendix: training data}

\begin{figure*}[h!]
    \centering
    \includegraphics[width=1\linewidth]{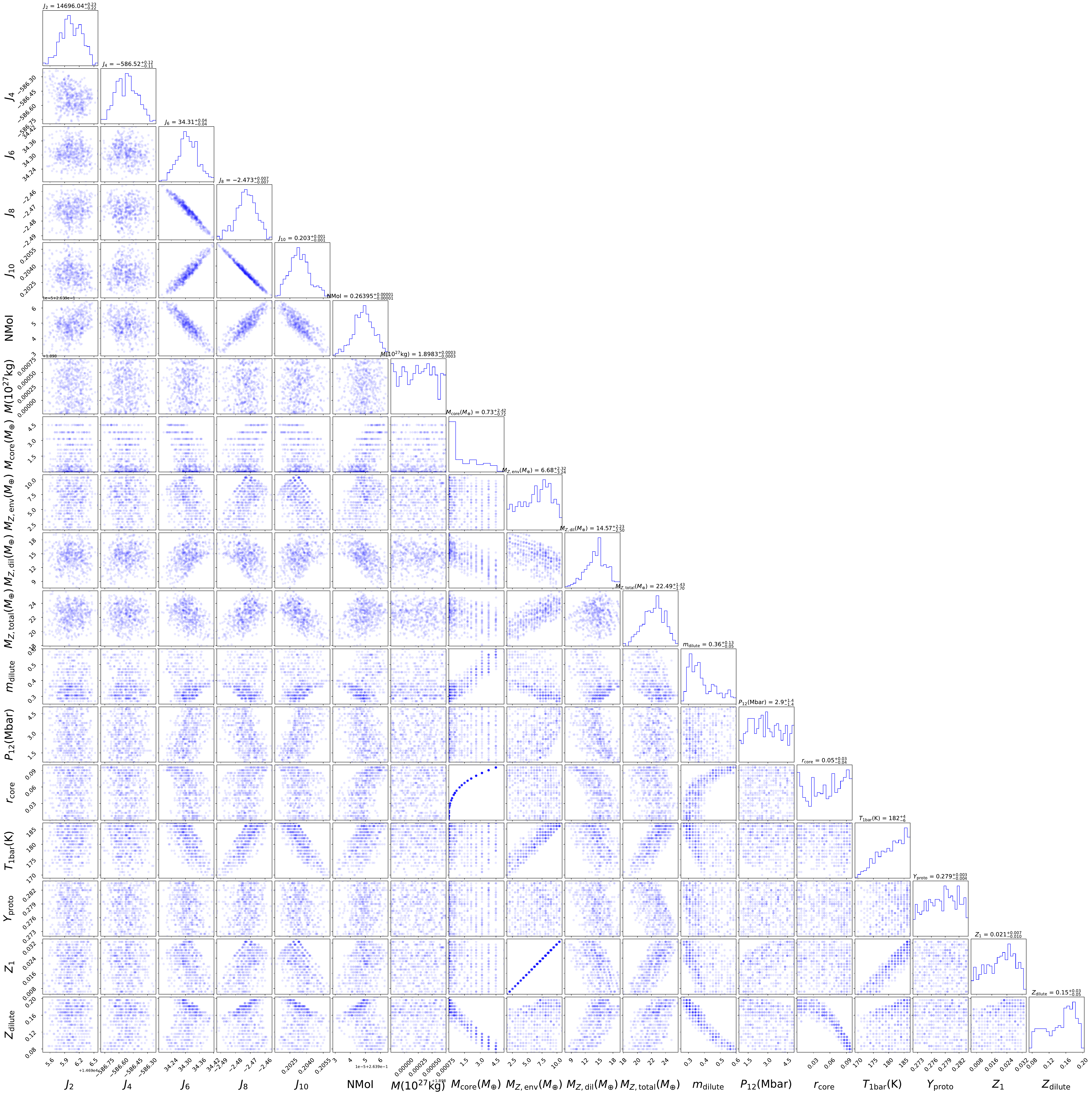}
    \caption{Resulting pair-wise relations and histograms of the observables and the interior parameters for the plausible models. The gravitational moments are scaled by $10^{6}$. The parameters for the distribution of the mass of heavy elements ($M_Z$) are defined the same as in \cite{Howard2023a}. Each point represents a specific interior model. Error bars are not shown for better readability.}
    \label{fig: Corner_plot}
\end{figure*}

Figure \ref{fig: Corner_plot} shows the pair-wise relations of the sample of 491 plausible interior models.

\twocolumn

\end{appendix}

\end{document}